\documentclass[12pt,letterpaper]{article}
\usepackage[a4paper, total={7in, 10in}]{geometry}

\usepackage{graphicx}
\usepackage{helvet}
\usepackage{authblk}
\usepackage{hyperref}
\usepackage{amsmath} 
\usepackage{amssymb} 
\usepackage{mathtools}
\usepackage{orcidlink} 
\usepackage[dvipsnames]{xcolor}
\usepackage{soul}
\usepackage[super,comma,sort&compress]  
   {natbib}
\usepackage[right]{lineno} \linenumbers

\makeatletter
\renewcommand{\maketitle}{\bgroup\setlength{\parindent}{0pt}
\begin{flushleft}
  \textbf{\@title}
  
  \@author
\end{flushleft}\egroup}
\makeatother

\title{Topological Localisation in Time from PT Symmetry}
\date{}

\author[1,\orcidlink{0009-0007-0655-0854}]{Tom Sheppard}
\author[1]{C. B. B. Camacho}
\author[2]{Sebastian Weidemann}
\author[2]{Alexander Szameit}
\author[2]{Joshua Feis}
\author[3]{Frank Schindler}
\author[1,*,$\dagger$]{Hannah M. Price}

\affil[1]{School of Physics and Astronomy, University of Birmingham, Birmingham B15 2TT, United Kingdom}
\affil[2]{Institute of Physics, University of Rostock, Rostock 18051, Germany}
\affil[3]{Blackett Laboratory, Imperial College London, London SW7 2AZ, United Kingdom}

\affil[*]{Lead Contact}
\affil[$\dagger$]{Correspondence: h.price.2@bham.ac.uk}

\begin{document}

\maketitle

\section*{SUMMARY}

Time has entered the domain of topological phases in the field of non-Hermitian physics. Previous studies have relied on periodic modulation in time to make an intuitive connection to established spatial topological invariants, albeit with energy and momentum exchanged. This connection has revealed the potential for topological interface states along the time axis, analogous to those in spatial models. In this work, we uncover a theoretical framework describing such topological interface states along the time axis, with no underlying connection to spatial models nor need for periodic driving. This new framework uncovers that this phenomenon -- the robust localisation of waves at an interface -- appears in \emph{every} system that has parity-time symmetry and two coupled modes or bands, regardless of its spatial dimensionality. The topological nature of this localisation is understood by the identification of certain topological phases that are specific to parity-time-symmetric models of two coupled modes. Our theoretical framework can be applied to all existing experimental observations, notably including photonic time crystals, and serves as a foundation for future experiments in areas in which the topological localisation of waves in time has yet to be studied.


\section*{KEYWORDS}


non-Hermitian physics, topology, time-varying metamaterials, photonic time crystals, PT symmetry 

\section*{INTRODUCTION}

The study of waves under the influence of externally imposed time variation has seen a surge of interest in recent years \cite{ yuan2022temporal, galiffi2022photonics, sharabi2022spatiotemporal, lustig2023photonic, pope20242024}. This has been accompanied by a range of new effects such as time-reflection and time-refraction at a temporal interface \cite{morgenthaler2003velocity,galiffi2022photonics, moussa2023observation, segal2026subcycletimerefractionopticalfrequencies}, frequency conversion in the absence of non-linear effects \cite{amra2024linear}, and non-reciprocal optical devices without the need for magnetic materials \cite{sounas2017non}. A key further example is broadband gain in periodically-driven photonic media. Rapid modulation of material parameters enable waves to draw energy from the external drive, creating an effective non-Hermiticity and thereby producing dispersions with momentum gaps -- extended ranges of momenta that feature energies with a non-zero imaginary part\cite{galiffi2022photonics, lustig2023photonic, pope20242024, xiong2025observation}.

Such a time modulation provides one pathway, amongst many others, to the realisation of non-Hermitian photonics which has also been the subject of intense study over the past decade \cite{ruter2010observation, zhu2014pt, peng2014parity, feng2017non, el2018non}. Notable phenomena include enhanced sensing \cite{chen2017exceptional, hodaei2017enhanced}, unidirectional invisibility \cite{liu2018unidirectional}, and non-Hermitian lasers \cite{hodaei2014parity}. A theoretical cornerstone of this area is the concept of parity-time-reversal ($\mathcal{PT}$) symmetry \cite{bender1998real, Ruschhaupt2005physical, el2007theory} where gain and loss, either inherent or effective, are required to be distributed symmetrically in space. Under this symmetry, modes may counter-intuitively have purely real energies or appear in complex-conjugated pairs. The regime of non-real energies leads to amplification and decay of waves and will be the focus of this work.

In spatially-modulated media, energy gaps arise in the dispersion and are well known to have a topological character \cite{haldane2008possible, wang2009observation, lu2014topological, yang2015topological, ozawa2019topological}. Lustig et al. \cite{lustig2018topological} were the first to consider if momentum gaps could possess non-trivial topology in the context of a binary photonic time crystal. If one exchanges space and time, the photonic material switches from a temporal crystal to a one-dimensional spatial crystal, where the Zak phases of energy bands can be calculated~\cite{zak1989berry, xiao2014surface}. Following this analogy, Lustig et al. proposed the Zak phase of a momentum band as the building block defining topological phases of momentum-gapped mediums. A hallmark of the topology of spatial crystals is the bulk-boundary correspondence -- the existence of bound states at the interface between two topologically distinct materials. The same authors\cite{lustig2018topological} also displayed simulation results showing that spatial bound states can have a temporal analogue -- a localisation (peaking) of the wave amplitude was observed at the temporal interface of two distinct mediums. Following these results, subsequent works \cite{lin2024temporally, zhu2024characterizing, ren2025observation, yang2025topologicallyprotectededgestates, feis2025space, xiong2025observation, tong2025observation, litopological2025} have carried out further investigations of temporally-localised waves arising due to momentum gaps, notably including experimental observations in fibre-optic based photonic quantum walks \cite{ren2025observation, feis2025space}, in an acoustic metamaterial \cite{tong2025observation}, in a microwave transmission-line photonic time crystal\cite{xiong2025observation}, and an active mechanical lattice\cite{litopological2025}.

Inspired by the increasing evidence of time localisation with a topological origin, we uncover the first rigorous framework that identifies localisation of waves at a temporal interface as a topological wave phenomenon across a broad class of models. The class is described by three properties: (1) the Schrödinger equation describes effective wave dynamics, (2) $\mathcal{PT}$ symmetry is present, and (3) the model describes two coupled modes or bands, with any number of spatial dimensions. At the centre of our approach is the identification of certain topological phases in a system of two parity-time-symmetric coupled modes. The existence of these phases is a consequence of the $\mathbb{Z}_2$ topological classification of imaginary-line-gapped zero-dimensional models in the non-Hermitian AI class \cite{kawabata2019symmetry, kitaev2001unpaired}; the present work identifies a physical consequence of this non-trivial classification for the case of two coupled modes or bands. Moreover, this notion of topology does not require space, or equivalently momentum, as an ingredient meaning the only dimension it is linked to is time. 

\section*{RESULTS}

\subsection*{Topological Localisation in Time in a Zero-Dimensional Model}

\begin{figure}[h]
    \centering
    \includegraphics{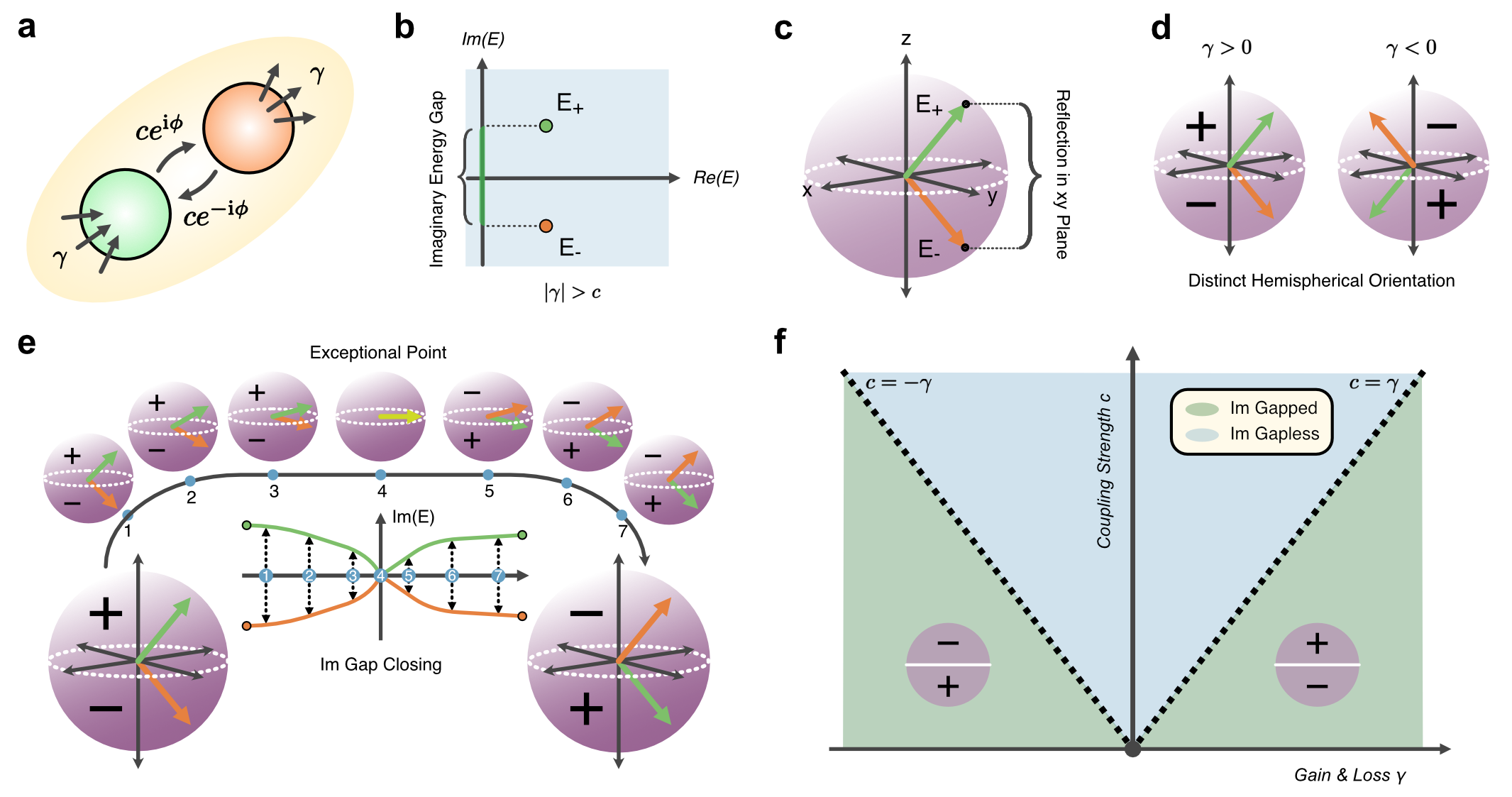}
    \caption{ Topological phases of two parity-time-symmetric coupled modes. \textbf{a} Two parity-time-symmetric coupled modes: Two resonators are coupled by a hermitian complex coupling $ce^{i\phi}$, with one site experiencing gain and the other loss both of which are of strength $\gamma$. \textbf{b} Imaginary gap: For $|\gamma| > c$, the spectrum features a separation in the imaginary part of the energy. \textbf{c} Reflection symmetry of eigenvectors: If the model has an imaginary gap, $\mathcal{PT}$ symmetry enforces that eigenvectors are related by a reflection in the xy-plane of the Bloch sphere. \textbf{d} Distinct hemispherical orientation of eigenvectors: The reflection symmetry of eigenvectors implies there are two distinct hemispherical orientations: either the eigenvector of $E_+$ lies in the +z or the -z hemisphere. These choices are determined by $\text{sign}(\gamma)$. \textbf{e} Topological distinction of hemispherical orientations: The hemispherical orientations of eigenvectors, enforced by the reflection symmetry, can only be continuously connected by closing the imaginary gap. \textbf{f} Topological phase diagram: For parameter values where the model is imaginary-gapped ($|\gamma| > c$), two distinct phases are identified by $\gamma > 0$ and $\gamma < 0$, corresponding to distinct hemispherical orientations of eigenvectors.}
    \label{fig:1}
\end{figure}

Consider a parity-time-symmetric model of two coupled modes. The state of the system will be described by a two-component complex vector $\psi$, representing the amplitude and phase of each mode. Dynamics of the system are determined by the Hamiltonian $H$ and follow from the Schrödinger equation $\mathrm{i}\dot{\psi} = H\psi$. The Hamiltonian is given by
\begin{equation}
    H = \begin{pmatrix}
        i\gamma & c e^{\mathrm{i}\phi} \\
        c e^{-\mathrm{i}\phi} & -i\gamma
    \end{pmatrix},
\end{equation}
illustrated schematically in Figure \ref{fig:1}a. Here $c$ and $\gamma$ are real parameters with $c \geq 0$, and $\phi \in [0,2\pi]$. The Hamiltonian satisfies $[H, \sigma_x \mathcal{K}] = 0$, where $\sigma_x$ is the Pauli x matrix and $\mathcal{K}$ is the complex conjugation operation \cite{Uhlmann_2016}; this is the statement of parity-time-reversal symmetry with $\mathcal{PT} = \sigma_x \mathcal{K}$. Whilst we have chosen to describe the system physically as coupled modes, observed for example in a pair of optical waveguides\cite{ruter2010observation}, we note that the same mathematical model appears in many different physical contexts\cite{klauck2019observation, martello2023coexistence, wu2019observation, li2019observation, ding2021experimental}. The energies $E_\pm$ are given by $E_\pm =\pm \sqrt{c^2 - \gamma^2}$ such that when $|\gamma| > c$, $E_\pm$ have a non-zero imaginary part with opposite sign. We describe such a spectrum as imaginary-gapped, as illustrated in Figure \ref{fig:1}b. Parity-time-reversal symmetry, in the case of an imaginary-gapped spectrum, mandates that the eigenvectors $v_\pm$ of energies $E_\pm$ are related by the $\mathcal{PT}$ operator. This is best understood when $v_\pm$ are placed on the Bloch sphere, as displayed in Figure \ref{fig:1}c, where the $\mathcal{PT}$ operator corresponds to a reflection in the xy-plane [S1]. Such a symmetry mandates that there are two possible \emph{hemispherical orientations} of eigenvectors: either the eigenvector of $E_+$ resides in the +z hemisphere or the -z hemisphere, with eigenvector belonging to $E_-$ lying in the respective other hemisphere. These orientations are identified by $\text{sign}(\gamma)$ and are shown in Figure \ref{fig:1}d.

Motivated by the search for topological phases, we consider continuous variation between Hamiltonians of differing eigenvector orientation, i.e. the growing ($E_+$) eigenvectors of each Hamiltonian are located in different hemispheres. As illustrated in Figure \ref{fig:1}e, to deform continuously between Hamiltonians of differing eigenvector orientation requires the imaginary gap to close at least once. Conversely, Hamiltonians of the same eigenvector orientation may be continuously connected whilst keeping the gap open. Two gapped Hamiltonians are topologically equivalent if and only if they can be continuously connected whilst maintaining the gap and symmetry \cite{kawabata2019symmetry}. Thus, we conclude that $\mathcal{PT}$ symmetry enables two distinct imaginary-gap topological phases. Further details on these phases are given in section S1. The resulting phase diagram is shown in Figure \ref{fig:1}f. 

Since zero-dimensional systems are always at an atomic limit, identifying a physically motivated \emph{topologically trivial} reference phase is not possible. Instead, we may look for observables that are sensitive to topological distinctions between the two different phases. To this end, we consider a \emph{temporal-topological interface} at time $t_0$ defined as evolution under a time-dependent Hamiltonian $H(t)$ such that for $t < t_0$, $H(t)$ belongs to a single gapped topological phase (meaning the eigenvector associated with $E_+$ remains in just one hemisphere), and for $t > t_0$, $H(t)$ belongs to the other phase. We stress that beyond this constraint the time dependence is arbitrary, and includes complex time-dependent Hamiltonian dynamics where in general $[H(t_1), H(t_2)] \neq 0$. Despite this complexity, we prove that the topological distinction between the two phases determines the characteristic behaviour of the wave intensity, computed as the vector-norm $|\psi(t)|^2 = \psi(t)^\dagger\psi(t)$, along the t-axis. This behaviour is a localisation (peaking) of the intensity about the temporal interface from any initial state that has largest overlap\footnote{By overlap, we refer to the biorthogonal overlap. Further details on this may be found in sections S1 and S2.} with the initial growing eigenvector. The rigorous proof of this behaviour is provided in complete technical detail within section S1. It is important to note that we use the term \emph{localisation} purely as a description of the intensity curve along the time axis; causality dictates that the intensity profile prior to the temporal interface is completely independent of that which comes after.

This topological localisation can be understood in qualitative terms in the following way: firstly, if the system is initialised such that it lies in the hemisphere containing the growing eigenvector, the intensity must increase at all times prior to $t_0$. This is irrespective of the time-variation of the Hamiltonian as the continual increase is a protected feature of the dynamics within a single topological phase. This can be understood by considering that if at some time $t < t_0$ the state of the system lies in the same hemisphere as the growing eigenvector, it must have largest overlap with that vector. At a short time $\delta t$ later, the intensity has increased, as a result of this larger overlap, and the state of the system remains in the same hemisphere as it was a time $\delta t$ earlier. Remaining in the same hemisphere can be intuitively understood from the fact that any component that the state has in the direction of the growing eigenvector at time $t$ can only be amplified across the time interval $\delta t$. Moreover, by virtue of no topological transition occurring prior to $t_0$, the growing eigenvector at time $t + \delta t$ must be in the same hemisphere as that at time $t$, and so this mechanism of intensity increase repeats. In short, the state of the system is trapped within the hemisphere in which it began, and the intensity grows continually as a result. 

Following the interface at $t_0$, the switch in topological phase guarantees that a decrease in the intensity is observed. This is understood qualitatively by noting that since the topological phase has changed, the hemisphere containing the growing and the hemisphere containing the decaying eigenvectors have exchanged. Immediately after $t_0$, the state of the system resides in the same hemisphere as the decaying eigenvector by virtue of its continuity across the temporal interface. Consequently, it has largest overlap with that vector. At a time $\delta t $ later, the intensity has decreased as a result of this larger overlap. However, the state of the system is not guaranteed to remain in the same hemisphere as it was $\delta t$ earlier. This is due to the fact that, on the Bloch sphere, the state of the system is generically attracted towards the location of the growing eigenvector. In other words, even a very small component of the state in the direction of the growing eigenvector will increase exponentially with time. Consequently, in the general case it takes a finite time $t_\mathrm{r}$ (which we call the regrowth time) after $t_0$ for the state to reach and cross the equator into the hemisphere where growing eigenvectors are contained. After this point, the situation is identical to that of times prior to $t_0$ and the intensity is observed to continually increase. To put it concisely, the intensity falls after the switch in topological phase for as long as the state remains in the hemisphere in which decaying eigenvectors are located.

\begin{figure}
    \centering
    \includegraphics{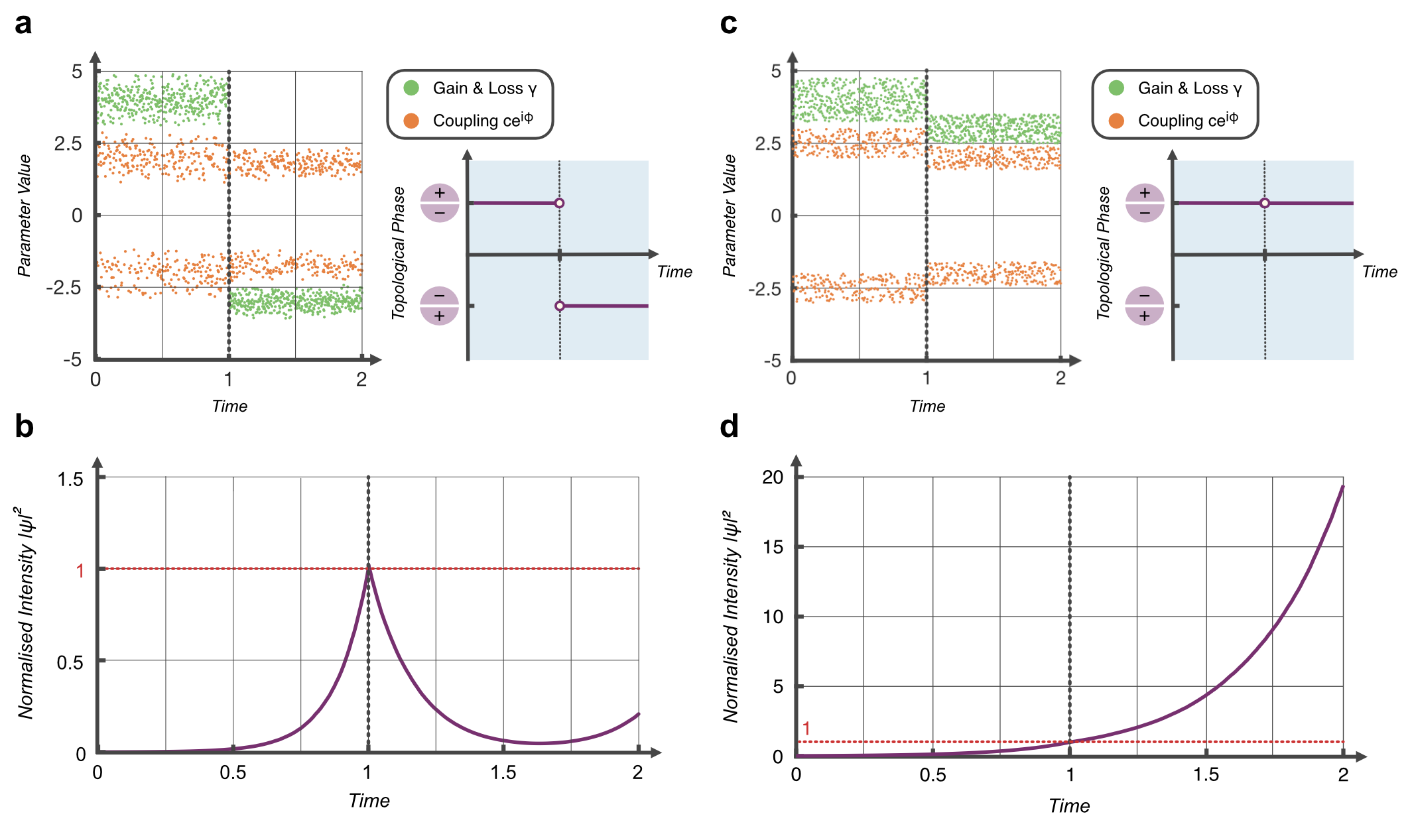}
    \caption{Robust localisation in time at a topological interface. \textbf{a} A topological time interface: The gain and loss $\gamma$ and coupling $ce^{i\phi}$ are chosen such that for $t < 1$ the model is in one topological phase (\emph{i.e.}, has a given hemispherical arrangement of eigenvectors as explained in Fig.~\ref{fig:1}), and for $t > 1$ it is in the other phase. Beyond this constraint, the time dependence of $\gamma$ and $c$ are generated by sampling a uniform distribution about a fixed mean value. Moreover, at each time step in the simulation, $\phi$ is randomly selected, with equal probability, to be either $0$ or $\pi$. This illustrates that the details of the time dependence, within each topological phase, may be arbitrary. \textbf{b} Localisation in time at the interface: The normalised wave intensity, computed as the norm $|\psi(t)|^2 / \left| \psi(1) \right|^2$, localises (peaks) in the temporal vicinity of the interface as a result of this topological change. At $t = 0$, the system is initialised in the instantaneous $E_+$ eigenvector and the value of the normalised intensity at the interface, which by definition is 1, is highlighted in red to demonstrate the clear peaking of the curve at that point. \textbf{c} A non-topological time interface: The parameters are chosen such that for both $t < 1$ and $t > 1$ the model is in a single topological phase, with differing average values. The parameters are randomly generated in a directly analogous manner to the topological case and the average magnitude of $\gamma$ and $c$ were chosen to be comparable to that case. \textbf{d} No localisation in time at the interface: The normalised wave intensity does not peak at the interface, as demonstrated by the red line indicating the value of the intensity at the interface. At $t = 0$, the system is initialised in the instantaneous $E_+$ eigenvector.}
    \label{fig:2}
\end{figure}

This time localisation effect is shown in Figures \ref{fig:2}a \& \ref{fig:2}b for a temporal-topological interface that features highly disordered parameter variation. Note that, as Figure \ref{fig:2} demonstrates, a periodic (or Floquet) drive is not required to observe this effect. Results from Floquet theory, notably the periodicity of quasi-energy, have been a central part of previous works that report temporally-localised waves with a topological origin; for example, this periodicity is necessary in order to define the Zak phase\cite{lustig2018topological} and chiral winding number\cite{yang2025topologicallyprotectededgestates} of a momentum band. Our results highlight that topological localisation in time is not an phenomenon exhibited only by Floquet systems, as may have been supposed based on previous work\cite{lustig2018topological, lin2024temporally, ren2025observation, yang2025topologicallyprotectededgestates, xiong2025observation, tong2025observation} that links the effect to momentum band invariants. In Figures \ref{fig:2}c \& \ref{fig:2}d, the behaviour of intensity across a non-topological interface, with parameter values comparable in scale to the topological interface, is shown; the intensity displays no localisation about the interface, increasing both before and after. For completeness, within section S3, figures S2 and S3 display the time-dependence of the mode amplitudes on individual sites for both the topological and non-topological interfaces simulated in Figure \ref{fig:2}.

In summary, parity-time symmetry present in a two-mode model facilitates non-trivial topology when the spectrum has an imaginary gap. The distinct topological phases are diagnosed by the \emph{hemispherical orientation} of eigenvectors, i.e. by which hemisphere the growing eigenvector is located in. This restriction to a single hemisphere in each phase enforces a localisation of the wave intensity at a temporal interface where this topology is mismatched. Whilst we have focused on the two-mode model whose parity-time symmetry operator is given by $\sigma_x\mathcal{K}$, any two-mode Hamiltonian that satisfies $[H, \mathcal{PT}] = 0$, where $\mathcal{PT}$ is an antiunitary operator squaring to the identity, has directly analogous non-trivial imaginary-gap topology and thereby features a robust intensity localisation effect [S1]. This invites the question as to whether localisation in time is a general phenomenon of imaginary-line-gapped models in the non-Hermitian AI class \cite{kawabata2019symmetry}, independent of the number of modes of the system. In section S6, we show that this is not the case: an explicit four-mode counterexample is provided in which a change in topological phase at a temporal interface does not lead to a localisation in the intensity. 

\subsection*{Topological Localisation in Time in Higher Spatial Dimensions}

We now consider a translationally-invariant spatial model with two local degrees of freedom that is $\mathcal{PT}$-symmetric, as an extension of the zero-dimensional case. As before, we assume that wave dynamics are accurately described by the Schrödinger equation. Examples include acoustic systems\cite{zhang2023observation}, exciton-polaritonic systems\cite{pickup2020synthetic}, and ultracold atoms\cite{takasu2020pt}. The translational symmetry implies that momentum $\mathbf{k}$ is \emph{conserved}. By \emph{conserved}, we mean more precisely that the dynamics of an initial excitation, itself described by some distribution of momenta, follow from the superposition of independent dynamics occurring at each $\mathbf{k}$. Dynamics of a single momentum excitation are determined by the momentum-space Hamiltonian $H(\mathbf{k})$ which, owing to the two local degrees of freedom, is a $2 \times 2$ matrix. Parity-time symmetry in a spatially-extended model manifests as $[H(\mathbf{k}), \mathcal{PT}] = 0$, where $\mathcal{PT}$ is an antiunitary operator squaring to the identity \cite{kawabata2019symmetry}. As a result, $H(\mathbf{k})$ may be interpreted as a zero-dimensional two-mode model with $\mathcal{PT}$ symmetry whose parameters depend on $\mathbf{k}$. This is precisely the model that was studied in detail in the previous section. A single momentum excitation at $\mathbf{k}_0$ effectively engineers two parity-time-symmetric coupled modes with Hamiltonian $H(\mathbf{k}_0)$, providing a path to observe topological localisation of the wave intensity in spatial dimensions greater than zero. This requires $H(\mathbf{k}_0)$ to have an imaginary gap both before and after a temporal interface, which itself switches between the two topological phases. Such a temporally-localised wave has a uniform profile along spatial axes, analogous to how a spatial-topological bound state has an intensity profile that is uniform in time. Any spatially-localised excitation to the system will contain a range of momenta. In section S4, we show in detail how this, rather than being a hindrance, may be exploited to engineer and observe a topologically localised wave at a temporal interface with an, in principle, arbitrary spatial profile. In simple terms, temporally-localised contributions from each momentum combine to give a temporally-localised total intensity. 

In order to guarantee temporally-localised contributions at each $\mathbf{k}$, the eigenvector associated with $E_+(\mathbf{k})$ must be dominantly excited prior to the temporal interface. This invites consideration of how the hemispherical orientation of eigenvectors at different momenta, each of which is fixed due to $\mathcal{PT}$ symmetry, are related. Consider a path $P$ in momentum space from $\mathbf{k}_1$ to $\mathbf{k}_2$, such that the dispersion $E(\mathbf{k})$ has an imaginary gap at $\mathbf{k}_1$ and $\mathbf{k}_2$, and remains gapped along the path $P$. This path represents a gap-preserving deformation of the zero-dimensional parity-time-symmetric two-mode model and therefore, must correspond to exploring a single topological phase of this model. It then follows that a path-connected domain $D$ of imaginary energy in momentum space has \emph{orientational order} of eigenvectors: every momentum in the domain $D$ has the same hemispherical orientation of eigenvectors. In other words, every $E_+$ eigenvector of momenta within $D$ lies in the same hemisphere, whilst every $E_-$ eigenvector lies within the opposing hemisphere. The orientational order of the domain $D$ is a conserved quantity under deformations of the band structure that do not completely close $D$. Therefore, this order may only be changed if the domain is completely closed and reopened, and thus is identified as a topological property of the domain. 

\begin{figure}
    \centering
    \includegraphics{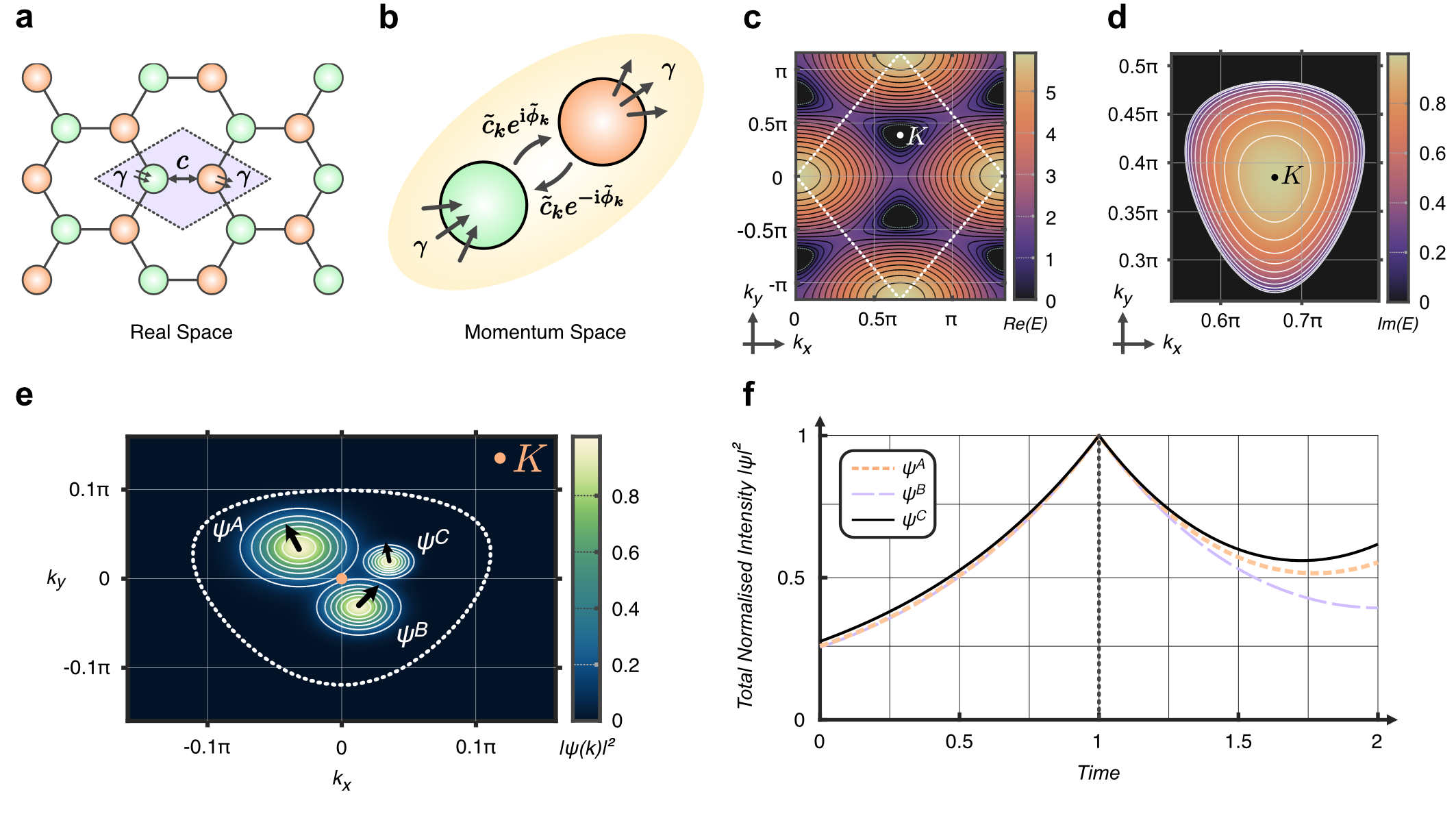}
    \caption{Topological localisation of light at a temporal interface in a model of photonic graphene. \textbf{a} The model in real space: Balanced gain \& loss $\gamma$ act on the distinct sublattices of the photonic honeycomb lattice. Neighbouring waveguides are coupled by the hopping parameter $c$. \textbf{b} The model in momentum space: In momentum space, the model acquires the interpretation as a series of decoupled two-mode models with parity-time symmetry, whose hopping parameters depend on momentum. \textbf{c} Real part of the bandstructure: The real part of the energy as a function of quasi-momentum for $c = 2$ and $\gamma = \pm1$. The two bands have equal and opposite real part, when non-zero, so we plot the modulus of the real part of each band. \textbf{d} Imaginary part of the bandstructure: The imaginary part of the energy as a function of quasi-momentum for $c = 2$ and $\gamma = \pm1$. The two bands have equal and opposite imaginary part, when non-zero, so we plot the modulus of the imaginary part of each band. \textbf{e} Wavepacket excitations in the momentum \emph{hole}: Three different wavepacket excitations within a single path-connected domain of imaginary energy are displayed. The colour indicates the relative intensity of excitation at each quasi-momentum. All three are designed such that they dominantly excite the $E_+$ eigenvector at each momentum, so that they do not immediately decay after excitation. Explicit forms of these initial states are given in section S4. \textbf{f} Topological localisation of the wave intensity at a temporal interface: The total normalised intensity, computed as $|\psi(t)|^2/|\psi(1)|^2$ with $\left| \psi \right|^2$ denoting the norm computed over the whole lattice, localises (peaks) in the temporal vicinity of the interface where $\gamma$ is switched from $1$ to $-1$, changing the eigenvector orientational order of the path-connected domain.}
    \label{fig:3}
\end{figure}

In one spatial dimension, or equivalently where momentum can be treated as a scalar, path-connected domains of non-real energy correspond to the widely studied momentum gap \cite{galiffi2022photonics, lustig2023photonic}. Thus, our results show that in the presence of $\mathcal{PT}$ symmetry and in two-band systems: (1) every momentum in the momentum gap can, in principle, host a topologically localised wave at a temporal interface and (2) momentum gaps feature \emph{eigenvector-orientational-order} as a topological property. In two spatial dimensions, the analogue of a momentum gap is a path-connected domain of imaginary energy in the $k_x$-$k_y$ plane, a \emph{momentum hole} if you like, whose topology may be switched at a temporal interface to facilitate localised waves. In Figure \ref{fig:3}, we numerically observe three examples of this localisation. The model simulated is based on photonic graphene with coupling $c$ and balanced gain and loss $\gamma$, following a recent waveguide implementation\cite{kremer2019demonstration}. This system is illustrated schematically in both real and momentum space in Figures \ref{fig:3}a \& \ref{fig:3}b. The bandstructure is displayed in Figures \ref{fig:3}c \& \ref{fig:3}d for $c = 2$ and $\gamma = \pm1$. For both choices of $\gamma$, we observe an identical region of imaginary energies whose topological phase switches as $\gamma$ is changed. Figure \ref{fig:3}f displays three intensity curves that are clearly localised about a topological interface, where $\gamma$ is switched from $+1$ to $-1$. Each curve corresponds to three different Gaussian initial states displayed in figure \ref{fig:3}e. Notably, these states contain a range of momenta, each of which contributes to the localisation of the total intensity. Further details on the model simulated and parameters chosen for these numerical results are given in section S4. 

In light of these results, we reflect that the introduction of spatial dimensions while retaining $\mathcal{PT}$ symmetry and two local degrees of freedom amounts to introducing the spatial profile of topologically localised waves as a new degree of freedom. Topology constrains temporal features, whilst leaving spatial features free to engineer. Moreover, the topological distinction of hemispherical orientations of eigenvectors, displayed in Figure \ref{fig:1}e, of the zero-dimensional model generalises to \emph{eigenvector-orientational-order} as a topological property of momentum gaps and their higher-dimensional analogues. 

\subsection*{Application to Photonic Time Crystals}

A class of systems that feature momentum gaps, currently under intense study, are photonic time crystals (PTCs)\cite{lustig2018topological, galiffi2022photonics, lustig2023photonic, lyubarov2022amplified}. These are spatially-homogeneous materials whose refractive index is periodically modulated in time.
It was Lustig \emph{et al.}\cite{lustig2018topological} that first considered the prospect of temporally-localised waves with a topological origin in their study of a binary PTC. Their approach involved defining momentum-gapped topological phases of the PTC in terms of the Zak phases of momentum bands. Following this, they then conjectured that an interface between PTCs of differing topology led to a temporally-localised wave appearing within the momentum gap and provided numerical evidence. In light of the momentum-gap topological phases described in the previous section, the question arises as to whether PTCs fall within the class of $\mathcal{PT}$-symmetric models in which these topological phases appear. If so, a further question arises as to whether these topological phases underpin the temporal localisation phenomena that have been observed both numerically\cite{lustig2018topological, lin2024temporally} and experimentally\cite{xiong2025observation} in PTCs. In this work, we prove that if a PTC has a time-reversal-symmetric modulation then both of these questions are answered in the affirmative. This proof can be found in complete technical detail within section S5. 

We can understand the mechanism behind topological localisation in time within a PTC in the following way: suppose that linearly-polarised light with momentum $k$ is incident on the PTC whose time period is $T$, whose periodically-modulated permittivity is $\epsilon(t)$, and whose permeability is $\mu$. As a result of spatial homogeneity, the momentum $k$ is conserved throughout wave propagation. The light may be considered at any given moment as the superposition of time-reflected (t-reflected) and time-refracted (t-refracted) waves [S5] with frequencies given by $\pm\omega(t) = \pm k/\sqrt{\epsilon(t) \mu}$. As a result, we may express the light's displacement field as

\begin{equation}
    D(z,t) = A(t) e^{-\mathrm{i} \omega(t) (t_0 + nT)} e^{\mathrm{i}(\omega(t)t + kz)} + B(t) e^{\mathrm{i} \omega(t) (t_0 + nT)} e^{\mathrm{i}(-\omega(t)t + kz)},
\end{equation}

where we have chosen z as the propagation direction. Here $A$ and $B$ are time-dependent complex numbers encoding the amplitude and phase of t-reflected and t-refracted waves, $n$ is the period number, and $t_0$ refers to the time about which $\epsilon(t)$ has time-reversal symmetry. By time-reversal symmetry, we mean more precisely that $\epsilon(t)$ satisfies $\epsilon(t_0 + \Delta t) = \epsilon(t_0 - \Delta t)$ for all time intervals $\Delta t$. Justification for the inclusion of the additional $e^{\pm\mathrm{i} \omega(t) (t_0 + nT)}$ phase factors is given in detail in section S5. We define the time-period transfer matrix\cite{yariv1982optical} $P(t)$ via

\begin{equation}
    \begin{pmatrix}
        A(t+T) \\
        B(t +T)
    \end{pmatrix} = P(t) \begin{pmatrix}
        A(t) \\
        B(t)
    \end{pmatrix}.
\end{equation}

$P(t)$ encodes the variation in amplitude and phase of the t-reflected and t-refracted waves across a period. It follows, as a consequence of the time-reversal symmetry of $\epsilon(t)$, that $P(t_0)$ satisfies $P(t_0)^* = P(t_0)^{-1}$ [S5]. This is nothing other than the statement of $\mathcal{PT}$ symmetry in the evolution operator sense\cite{mochizuki2016explicit}; $P(t_0)$ can be written as $\exp(-\mathrm{i}H_PT)$ where $H_P$ is a $\mathcal{PT}$-symmetric Hamiltonian satisfying $[H_P,\mathcal{K}] = 0$. Therefore, the change in the t-reflected and t-refracted waves that occurs from time $t_0$ to $t_0 + T$ can be mapped to the evolution of a parity-time-symmetric system of two coupled modes whose Hamiltonian is $H_P$. In other words, $A$ and $B$ change from period to period as if they were a pair of modes coupled in a $\mathcal{PT}$-symmetric fashion. $A(t)$ and $B(t)$ at times $t_0, t_0 + T, t_0 + 2T, \dots$ determine the strength of the light's displacement field according to the relation $|D|^2 = |A + B|^2$. We note that this does not translate into the intensity $|\psi|^2$ within the coupled-mode picture of this evolution. This is due to that fact that, within this picture, modes are conventionally treated\cite{haus1984waves} as being spatially-orthogonal meaning that the intensity would be given as $|\psi|^2 = |A|^2 + |B|^2$. Nevertheless, $|A+B|^2$ localises about a temporal-topological interface by precisely the same overlap mechanism as the coupled-mode intensity $|A|^2 + |B|^2$. Further details are provided in section S5.

To summarise, a PTC whose modulation is time-reversal-symmetric features wave propagation that, when examined stroboscopically, becomes completely equivalent to the dynamics of two parity-time-symmetric coupled modes. As a result, the momentum gaps of PTCs possess eigenvector-orientational-order as a topological property; see the previous section for a detailed discussion of these momentum-gap topological phases. The switching of this topology at a temporal interface then facilitates the stroboscopic localisation of light at that interface. Such a stroboscopic confinement of light is consistent with simulation evidence presented by Lustig \emph{et al}.\cite{lustig2018topological}, where light was observed to show clear localisation when observed at times $t_0, t_0 + T, t_0 + 2T, \dots$ and oscillate in its amplitude between those times. Moreover, as discussed in detail within the last section, every momentum in the momentum gap hosts a temporally-localised wave. A consequence of this property was recently seen in a transmission-line PTC engineered by Xiong \emph{et al.}\cite{xiong2025observation}; temporally-localised waves at five different momenta within the gap were observed.

\section*{DISCUSSION}

In conclusion, by identifying the imaginary-gap topological phases of two parity-time symmetric coupled modes, we have shown that the topological localisation of waves at a temporal interface appears as a universal phenomenon across a broad class of systems. The class is described by three properties: (1) two local degrees of freedom, (2) $\mathcal{PT}$ symmetry, and (3) dynamics described by the Schrödinger equation, irrespective of spatial dimensionality. This finding runs counter to what one would have inferred from previous literature\cite{lustig2018topological, lin2024temporally, zhu2024characterizing, ren2025observation, yang2025topologicallyprotectededgestates, feis2025space, xiong2025observation, tong2025observation, litopological2025} reporting temporally-localised waves with a topological origin which have exclusively examined one-dimensional systems. Moreover, these works proposed topological invariants underpinning the phenomenon that are written in terms of geometric properties of eigenmodes with real energies, without natural generalisations to systems with a different number of spatial dimensions. Many of these invariants are directly analogous, albeit with energy and momentum exchanged, to the invariants found in the celebrated topological band theory\cite{bansil2016colloquium} which underpin localised waves at spatial interfaces. However, this approach overlooks a key difference between spatial and temporal interfaces. In the temporal case, one always has \emph{direct access} to the details of a temporally-localised wave in the Hamiltonian of the system; this wave will be composed of amplifying and dissipative modes before and after the interface. In the spatial case however, details of the spatially-localised wave are not directly accessible from the bulk Hamiltonian either side of the interface, only the existence of the wave is able to be inferred by examining subtle global properties. As a result of this distinction, the topological features of temporally-localised waves are much more naturally captured by direct analysis of the growing and decaying modes of the system which is the starting point to obtain the results of the present work. This new approach linking topology and time invites further theoretical investigation of zero-dimensional non-Hermitian topology\cite{kawabata2019symmetry} and its consequences for the behaviour of waves along the time axis.

To the best of our knowledge, all existing experiments reporting temporally-localised waves with a topological origin have been made within the class of $\mathcal{PT}$-symmetric systems to which our results apply. These are observations made in fibre-optic based photonic quantum walks\cite{ren2025observation, feis2025space}, an acoustic metamaterial\cite{tong2025observation}, an active mechanical lattice\cite{litopological2025}, and a transmission-line photonic time crystal (PTC)\cite{xiong2025observation}. Our results therefore bring together these observations under a single theoretical framework and demonstrate that their topological nature can be understood by a simple geometric argument on the Bloch sphere. Moving beyond existing observations, our results provide a theoretical foundation for future experiments seeking to observe this phenomenon in physical platforms beyond those studied so far. Such experimental efforts would be aided in large part by the considerable attention that $\mathcal{PT}$ symmetry has received within the literature\cite{bender1998real, ruter2010observation, zhu2014pt, peng2014parity, el2018non, liu2018unidirectional, hodaei2014parity, klauck2019observation, martello2023coexistence, wu2019observation, li2019observation, ding2021experimental, kremer2019demonstration}. A system of particular interest\cite{lyubarov2022amplified} is a PTC in the optical regime, which has yet to be realised experimentally. Despite the substantial challenges involved, Segal and coauthors have recently made an important advance by realising a sub-wave-cycle permittivity change in a sample of doped Cadmium Oxide\cite{segal2026subcycletimerefractionopticalfrequencies}. As advances continue to bring optical PTCs within experimental reach, our theory may serve as a valuable tool for future realisations of topological localisation in time at optical frequencies.

\section*{Methods}

Further details regarding the methods can be found in the Supplemental Information.

\section*{RESOURCE AVAILABILITY}


\subsection*{Lead contact}


Requests for further information and resources should be directed to and will be fulfilled by the lead contact, Hannah M. Price (h.price.2@bham.ac.uk).

\subsection*{Materials availability}


This study did not generate new materials.

\subsection*{Data and code availability}


This study did not generate data. Code used to run numerical simulations in this study is available from the lead contact upon request.

\section*{ACKNOWLEDGMENTS}


T.S., C.C., and H.M.P. are supported by the Royal Society via grants UF160112, URF\textbackslash R\textbackslash221004, RGF\textbackslash E A\textbackslash180121 and RGF\textbackslash R1\textbackslash180071, and by the Engineering and Physical Sciences Research Council (grant no. EP/W016141/1 \& EP\textbackslash Y01510X\textbackslash1). A. S. acknowledges funding from the Deutsche Forschungsgemeinschaft (DFG, German Research Foundation) through SFB 1477 “Light-Matter Interactions at Interfaces” (project no. 441234705), IRTG 2676/1-2023 ‘Imaging of Quantum Systems’, (project no. 437567992) and grants SZ 276/9-2, SZ 276/19-1, SZ 276/20-1, SZ 276/21-1, SZ 276/27-1. A. S. also acknowledges funding from the FET Open Grant EPIQUS (grant no. 899368) within the framework of the European H2020 programme for Excellent Science, as well as from the Krupp von Bohlen and Halbach foundation. J.F. acknowledges support by the Leverhulme Trust through a Study Abroad Studentship. F.S. was supported by a UKRI Future Leaders Fellowship MR/Y017331/1. T.S. would also like to thank J. M. F. Gunn, Nick Jones, Oliver Ashfield, and Samuel Pickering for eye-opening discussions.

\section*{AUTHOR CONTRIBUTIONS}


T.S. developed the ideas and results present in this work, in discussion with the other authors. T.S. drafted the manuscript and all authors edited the paper. H.M.P. supervised this project.

\section*{DECLARATION OF INTERESTS}


The authors declare no competing interests.

\section*{DECLARATION OF GENERATIVE AI AND AI-ASSISTED TECHNOLOGIES IN THE WRITING PROCESS}

During the preparation of this work, the authors used ChatGPT in order to proofread written material. After using this tool, the authors reviewed and edited the content as needed and take full responsibility for the content of the publication.

\section*{SUPPLEMENTAL INFORMATION INDEX}




Document S1. Sections S1-S6, Figures S1–S5.

\newpage

\newpage


\bibliography{references}

\newpage



\end{document}